\documentclass{article}

    \PassOptionsToPackage{numbers, compress}{natbib}

 \usepackage[dblblindworkshop, final]{neurips_2025}
\usepackage{graphicx}
\workshoptitle{Foundation Models for the Brain and Body}

\usepackage[utf8]{inputenc} 
\usepackage[T1]{fontenc}    
\usepackage{hyperref}       
\usepackage{url}            
\usepackage{booktabs}       
\usepackage{amsfonts}       
\usepackage{nicefrac}       
\usepackage{microtype}      
\usepackage{xcolor}         
\usepackage{subcaption}
\usepackage{graphicx} 
 \usepackage{amsmath}
 
\title{Decoding Predictive Inference in Visual Language Processing via Spatiotemporal Neural Coherence}

\author{%
   Sean C.~Borneman\\
  Department of Physics\\
  Carnegie-Mellon University\\
  Pittsburgh, PA 15213 \\
  \texttt{sbornema@andrew.cmu.edu} \\
  \And
  Julia Krebs \\
  Department of Linguistics\\
  University of Salzburg \\
  Salzburg, Austria 5020\\
  \texttt{julia.krebs@plus.ac.at} \\
  \AND
  Ronnie Wilbur \\
  Department of Linguistics\\
  Purdue University \\
  West Lafayette, IN  47904\\
  \texttt{wilbur@purdue.edu} \\
  \And
  Evie A.~Malaia\\
  Department of Communicative Disorders\\
  University of Alabama\\
  Tuscaloosa, AL 35406\\
  \texttt{eamalaia@ua.edu} \\
}

\begin{document}

\maketitle

\begin{abstract}
Human language processing relies on the brain's capacity for predictive inference. We present a machine learning framework for decoding neural (EEG) responses to dynamic visual language stimuli in Deaf signers. Using coherence between neural signals and optical flow-derived motion features, we construct spatiotemporal representations of predictive neural dynamics. Through entropy-based feature selection, we identify frequency-specific neural signatures that differentiate interpretable linguistic input from linguistically disrupted (time-reversed) stimuli. Our results reveal distributed left-hemispheric and frontal low-frequency coherence as key features in language comprehension, with experience-dependent neural signatures correlating with age. This work demonstrates a novel multimodal approach for probing experience-driven generative models of perception in the brain.
\end{abstract}

\section{Introduction}

The brain implements hierarchical generative models that minimize prediction errors by aligning top-down predictions with sensory evidence \cite{friston2008hierarchical, auksztulewicz2015attentional}. According to predictive coding theory, neural networks maintain internal models of environmental structure, continuously updating predictions to minimize free energy—an information-theoretic measure of surprise \cite{friston2018does}. While predictive coding is established in auditory speech comprehension \cite{giraud2012cortical}, its mechanisms in visual language remain less understood.

Sign languages offer a compelling testbed for visual predictive coding because they involve rich spatiotemporal dynamics requiring real-time hierarchical inference. Signed stimuli exhibit greater Shannon entropy and information density than non-linguistic actions \cite{malaia2016a, borneman2018a}, reflecting structured linguistic content embedded in continuous motion. Deaf signers demonstrate enhanced sensitivity to dynamic visual patterns underlying comprehension \cite{malaia2021low}, engaging language-network regions when viewing signs, whereas non-signers rely predominantly on visual processing areas \cite{malaia2012event}. This experience-dependent neural specialization suggests that lifelong exposure shapes internal generative models for visual language.

\textbf{Problem Formulation}: We formalize sign language comprehension as hierarchical temporal inference under uncertainty. A fluent signer's brain maintains multi-scale generative models predicting upcoming visual input based on internalized linguistic structure. Successful comprehension corresponds to minimizing prediction error across hierarchical levels: high-level expectations (syntactic/semantic transitions) constrain lower-level perceptual predictions (kinematic features), consistent with theories of hierarchical predictive coding.

\textbf{Approach}: We introduce a coherence-based multimodal fusion approach combining EEG and computer vision to probe hierarchical predictive coding in sign language. We recorded EEG from Deaf signers viewing normal versus time-reversed sign language videos (preserving motion energy but destroying temporal linguistic structure), extracted optical flow motion features, and computed frequency-resolved coherence between neural and visual signals. Using machine learning with entropy-based feature selection, we decode brain states and identify neural mechanisms underlying structured visual language processing.

\textbf{Contribution}: Our approach demonstrates that: (1) neural coherence features discriminate structured vs. unstructured visual language inputs, revealing frequency-specific oscillatory mechanisms consistent with predictive coding; (2) language experience modulates these dynamics, with greater exposure correlating with stronger low-frequency neural entrainment; (3) multimodal fusion enables reverse-engineering of hierarchical internal models from neural-stimulus coupling patterns.

\subsection{Related Work}

Predictive coding frames perception as Bayesian inference with hierarchical generative models \cite{ramstead2021neural}. Brain oscillations track temporal structure in sensory input \cite{ding2016a}, with low-frequency dynamics carrying high-level predictions about structured events. Recent studies demonstrate cortical tracking of biological motion: Shen et al. showed MEG signals entrain to hierarchical kinematic patterns with stronger low-frequency coherence for structured movements \cite{shen2023cortical}. In sign language, low-frequency neural entrainment is critical for comprehension \cite{malaia2021low}, similar to delta-band oscillations in speech parsing.

Multimodal fusion approaches combining neural data with stimulus features have proven effective for decoding cognitive states \cite{de2018decoding}. Our work extends this by fusing EEG with optical flow to quantify neural tracking of complex visual motion, using entropy-based methods to characterize linguistic complexity \cite{borneman2018a, solorio2020review}. This represents the first application of EEG-optical flow coherence to sign language comprehension with age-dependent analysis.

\section{Methods}

\subsection{Participants \& Stimuli}
24 Deaf native signers of Austrian Sign Language (ÖGS), ages 20-60s (M=42, SD=12.27), with normal vision and ÖGS as primary language. All participants acquired ÖGS from birth, ensuring age serves as a valid proxy for cumulative language exposure in this population. Study was approved by local ethics board with informed consent, with 30 Euro/hour remuneration for participants.
Prior to data collection, participants received comprehensive instructions in Austrian Sign Language (ÖGS) with German translation available. Instructions specified the experimental task: watch sign language videos and evaluate sentence acceptability using mouse responses. A practice session familiarized participants with procedures before the main experiment comprising ten 5-minute blocks with self-paced breaks between blocks. Experimenters remained present throughout to address questions during breaks.

40 signed sentences in ÖGS plus time-reversed versions created by temporal reversal. This manipulation preserves local motion statistics but destroys linguistic temporal predictability, creating visually matched but linguistically meaningless controls analogous to reversed speech. Participants performed attention judgments confirming structured vs. unstructured stimulus interpretability.

\subsection{EEG Recording \& Preprocessing}
26 electrodes (10/20 layout), 500Hz sampling, 1-100Hz band-pass filtering. Signals preprocessed with artifact removal, mastoid re-referencing, and epochs time-locked to video presentations. This provided robust neural activity measures across widespread cortical regions.

\subsection{Optical Flow \& Neural Coherence Analysis}

\textbf{Motion Feature Extraction}: Horn-Schunck algorithm computed motion vectors between consecutive video frames \cite{horn1981a}:
$$\mathbf{u}(x,y,t) = \arg\min_{\mathbf{u}} \left[ \|\nabla I \cdot \mathbf{u} + I_t\|^2 + \alpha^2 \|\nabla \mathbf{u}\|^2 \right]$$
where $I(x,y,t)$ is image intensity, $\mathbf{u} = (u_x, u_y)$ are velocity components, and $\alpha$ controls smoothness. We derived univariate motion signals by spatially aggregating flow magnitudes at ~30Hz:
$$M(t) = \frac{1}{N} \sum_{x,y} \|\mathbf{u}(x,y,t)\|$$
capturing overall motion dynamics while preserving temporal linguistic structure.

\textbf{EEG-Motion Coherence}: Prior to coherence computation, EEG and optical flow time-series were aligned and resampled to common 30Hz rate. Coherence at frequency $f$ was computed as:
$$C_{xy}(f) = \frac{|S_{xy}(f)|^2}{S_{xx}(f) S_{yy}(f)}$$
where $S_{xy}(f)$ is the cross-power spectral density between EEG channel $x$ and motion signal $y$, and $S_{xx}(f)$, $S_{yy}(f)$ are auto-power spectral densities. We focused on 0.5-12Hz range encompassing dominant temporal modulations in sign language. 

Electrodes were grouped into four regional clusters (Left/Right × Anterior/Posterior) motivated by known language lateralization differences \cite{malaia2021salience}. For each region and frequency bin, we obtained coherence magnitude and optimal temporal lag, resulting in spatiotemporal coherence profiles characterizing neural-stimulus coupling.

\subsection{Feature Selection \& Machine Learning}

The full coherence feature set yielded 496 features (magnitude/timing × 62 frequency bins × 4 regions) for 24 participants. We employed unsupervised feature selection based on information-theoretic criteria, Shannon entropy (1) and mutual information (2):

\begin{equation}
H(X) = -\sum_{i} p(x_i) \log p(x_i)
\end{equation}
\begin{equation}
I(X; Y) = H(X) - H(X|Y)
\end{equation}

Features maximizing mutual information $I(X; Y)$ with target variables were selected, ensuring interpretable dimensionality reduction while preserving predictive signal for both age-related changes and stimulus-specific neural entrainment.

\textbf{Modeling Pipelines:} Two complementary approaches: (A) Age-targeted regression identifying features predictive of experience, comparing performance across algorithms (Linear Regression, LASSO, Elastic Net, kNN, Random Forest, Gradient Boosting); (B) Stimulus-driven classification identifying features discriminating structured vs. unstructured input, refined via recursive feature elimination for age correlation analysis. Cross-validation ensured robust performance estimates.



Note that feature selection in pipeline B was optimized for stimulus discrimination (structured vs. unstructured), not age prediction. This approach avoids circularity in subsequent age-correlation analyses.

\section{Results}

\subsection{Age Prediction from Neural Coherence}

We predicted participant age from EEG-optical flow coherence features using multiple regression algorithms. Model performance quantified by mean squared error: $\text{MSE}(y, \hat{y}) = \frac{1}{n} \sum_{i=0}^{n-1} (y_i - \hat{y}_i)^2$.

\begin{figure}[ht]
  \centering
  \begin{subfigure}[b]{0.48\textwidth}
    \includegraphics[width=\linewidth]{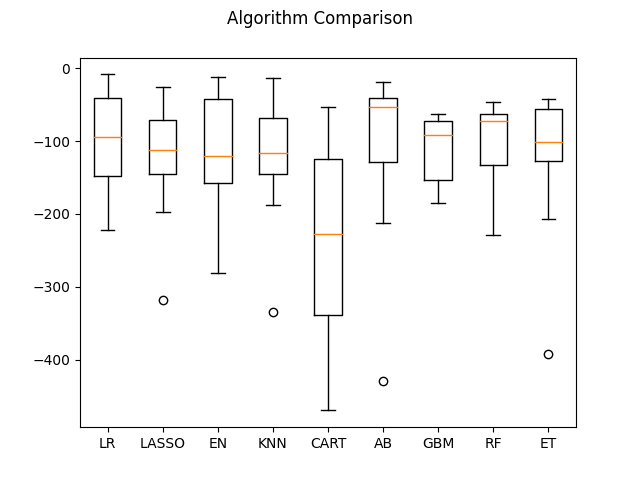}
    \caption{Full feature set.}
  \end{subfigure}
  \hfill
  \begin{subfigure}[b]{0.48\textwidth}
    \includegraphics[width=\linewidth]{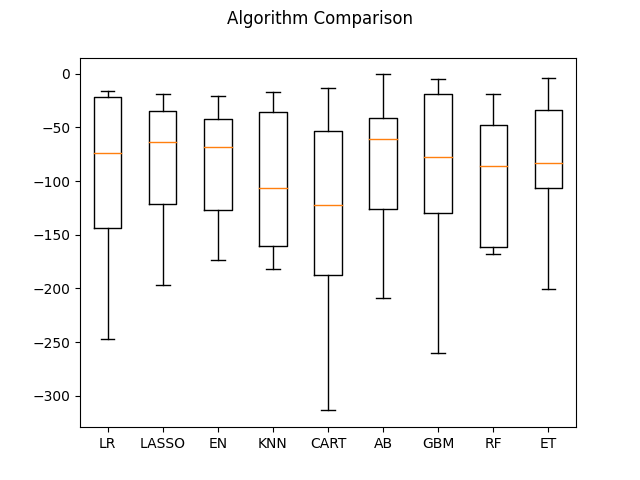}
    \caption{Top 10 entropy-selected features.}
  \end{subfigure}
  \caption{Age prediction performance. Boxplots show cross-validated MSE distributions (lower = better). Feature selection improved consistency for tree-based models while maintaining performance for regularized approaches. Baseline (mean age prediction): MSE = 151 years$^2$; best models achieved ~85-100 years$^2$ (RMSE ~9-10 years).}
  \label{fig:regression_models}
\end{figure}

With full features, best models achieved MSE ~85-100 years$^2$ (RMSE ~9-10 years), substantially outperforming baseline mean-age prediction (MSE = 151 years$^2$). LASSO, Elastic Net, and Gradient Boosting performed best. Using top 10 entropy-selected features maintained or improved performance, indicating most predictive signal captured by key features. Low-frequency timeshift features in frontal/posterior regions and high-frequency correlations in lateral areas emerged as strongest age predictors, suggesting experience modulates predictive neural timing across multiple temporal scales.

\subsection{Stimulus Classification Performance}

Neural coherence features achieved near-perfect classification of structured vs. unstructured input (cross-validated accuracy = 94.2 ± 3.1\%), demonstrating categorically distinct brain states during linguistic vs. non-linguistic visual processing. This supports predictive coding theory: structured input conforming to internal models elicits strong prediction signals, while violations result in breakdown of neural-stimulus alignment.

\subsection{Experience-Dependent Neural Signatures}

\begin{figure}[ht]
  \centering
  \begin{subfigure}[t]{0.48\textwidth}
    \includegraphics[width=\linewidth]{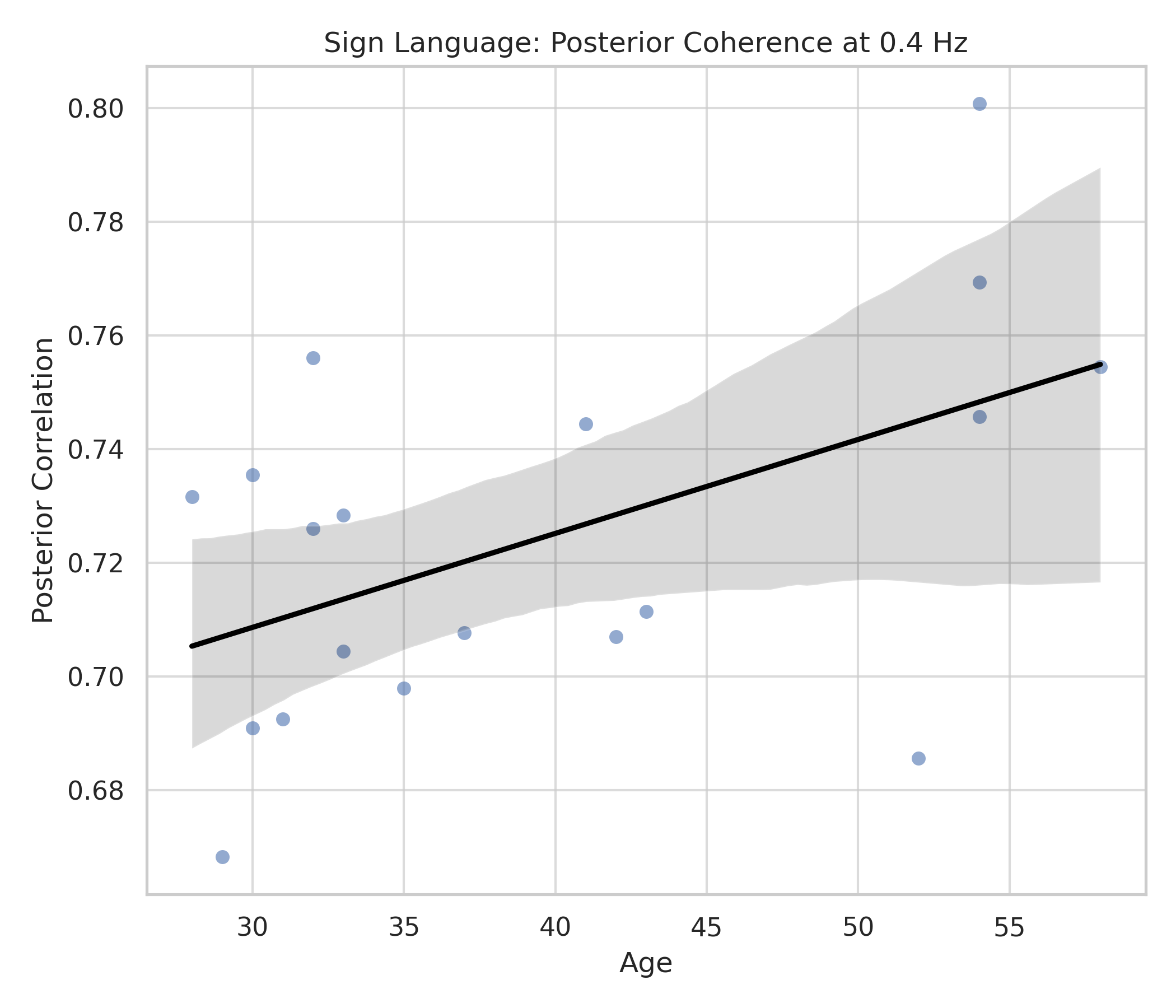}
    \caption{Posterior coherence (0.4Hz) during structured signing increases with age.}
  \end{subfigure}
  \hfill
  \begin{subfigure}[t]{0.48\textwidth}
    \includegraphics[width=\linewidth]{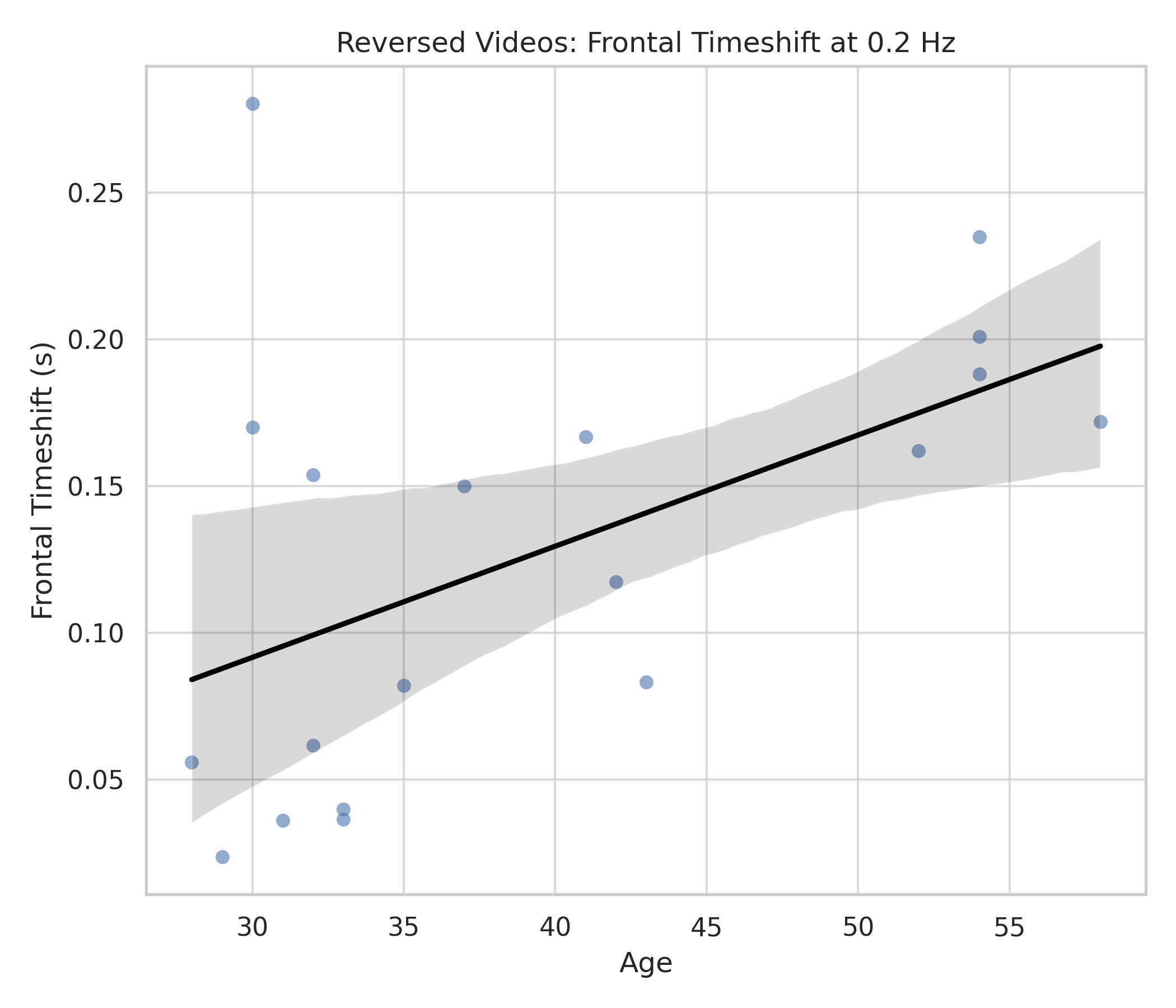}
    \caption{Frontal delays (0.2Hz) during reversed input increase with age.}
  \end{subfigure}
  \caption{Age-related changes in predictive dynamics. (a) Enhanced low-frequency coherence suggests improved higher-order predictions with experience. (b) Increased delays to unstructured input reflect greater reliance on learned generative models that incur larger prediction error costs when violated.}
  \label{fig:age_effects}
\end{figure}

Two specific coherence features showed significant age correlations, as depicted in Figure~\ref{fig:age_effects}. 
In structured sign language, posterior coherence at \(\sim 0.4\text{ Hz}\) increased with age (\(r \approx 0.51, p < 0.03\)), 
suggesting older signers exhibit stronger neural synchronization over 2--3~s intervals during linguistic processing. 
In reversed videos, frontal temporal lag at 0.2~Hz correlated with age (\(r \approx 0.52, p < 0.03\)), 
indicating older participants showed longer neural delays to unpredictable motion. 
These patterns indicate experience refines internal models: older signers show enhanced prediction over longer timescales 
for structured input but increased processing costs when predictions are violated.

\section{Discussion}

Our findings demonstrate how experience tunes hierarchical generative models in visual language processing. The coherence-based approach reveals that Deaf signers' brains exhibit categorically different states when viewing structured versus unstructured visual language, with measurable neural coupling at timescales of linguistic units. This aligns with predictive coding theory: structured input conforming to internal models elicits strong prediction signals, while violations result in breakdown of neural-stimulus alignment.

\textbf{Hierarchical Inference Mechanisms}: Age-related patterns suggest experience optimizes internal representations across temporal scales. Younger signers may rely primarily on local motion cues and short-range predictions, while older signers demonstrate neural evidence of integration over longer timescales corresponding to syntactic/pragmatic levels. The positive correlation between age and 0.4Hz coherence indicates experienced signers synchronize with multi-sign units, consistent with hierarchical inference where higher-level predictions involve slower neural dynamics \cite{blumenthal-dram2019a}.

\textbf{Computational Interpretation}: The trade-off between enhanced structured processing and increased costs for random input reflects optimization for typical linguistic environments. From a computational perspective, this supports viewing the brain as an adaptive generative model that improves inference under familiar conditions at the expense of flexibility with anomalous inputs—a hallmark of experience-dependent specialization.

\textbf{Free Energy Minimization}: Our results provide empirical evidence for free energy minimization in visual language. The coherence differences between structured and unstructured conditions can be interpreted as neural signatures of prediction error: reduced coherence to reversed stimuli reflects increased sensory surprise when internal models fail to predict input structure. The age-dependent increase in this effect suggests refined models generate stronger predictions (and thus larger prediction errors when violated).

\textbf{Methodological Innovation}: Our entropy-based feature selection identified interpretable neural signatures linked to specific frequency bands and brain regions. The multimodal fusion approach demonstrates how combining neural and stimulus-derived features can reveal meaningful brain-behavior relationships while maintaining interpretability through careful dimensionality reduction. This provides a generalizable framework for studying experience-driven neural adaptation in complex natural tasks.

\section{Limitations \& Future Work}

Key limitations include small sample size (N=24) requiring replication in larger cohorts, and population specificity (Austrian Sign Language users) necessitating cross-linguistic validation. While age serves as a valid proxy for cumulative language exposure in native signers who acquired ÖGS from birth, future studies could incorporate direct measures of language proficiency for finer-grained modeling. EEG provides excellent temporal but limited spatial resolution; future work combining with fMRI/MEG could improve anatomical precision of predictive coding localization. Additional controls beyond time-reversal (e.g., spatially scrambled or semantically violated signs) would strengthen claims about linguistic specificity. The current approach focuses on motion-based features; incorporating hand shape and spatial configuration could reveal complementary aspects of sign language neural processing.

\section{Broader Impact}

This work has implications for energy-efficient AI systems by demonstrating predictive strategies that treat predictable inputs as requiring minimal processing. Computer vision and sign language recognition systems could incorporate multi-timescale predictive mechanisms, focusing computational resources on unexpected movements while auto-completing predictable sequences. The experience-dependent optimization observed here suggests AI systems could similarly adapt their internal models based on domain-specific exposure. Neural coherence patterns could serve as biomarkers for healthy brain aging or cognitive assessment, given the systematic relationship between experience and predictive neural signatures. The approach provides tools for studying neural plasticity and specialization across diverse populations and sensory modalities.

\bibliographystyle{unsrtnat}
\bibliography{refs_cleaned}

\end{document}